\documentstyle[preprint,aps,epsf]{revtex}

\newcommand{\beq}{\begin{eqnarray}}
\newcommand{\eeq}{\end{eqnarray}}

\newcommand{\btem}{\bibitem}

\tightenlines

\begin{document}

%\preprint{UTHEP-362, June. 1997}

\draft

\title{Variational Perturbation Approach \\ to Resonance-State
 Wave Functions}

\author{T. Tanaka}

\address{Institute of Physics, University of Tsukuba,
 Tsukuba, Ibaraki 305, Japan
%\\ E-mail: jstthp@nucl.ph.tsukuba.ac.jp}
}

%\date{\today}
 
\maketitle

\begin{abstract}
The variational perturbation theory for wave functions, which has
 been shown to work well for bound states
 of the anharmonic oscillator, is applied to resonance states
% This is achieved by allowing the variational parameter
% to take complex values. We apply this generalized method to
 of the anharmonic oscillator with negative coupling constant.
 We obtain uniformly accurate wave functions
 starting from the bound states.
% wave functions.

\end{abstract}

\bigskip
%\pacs{PACS numbers: 02.30.Mv; 11.15.Tk\\
%{\it Keywords}: Variational perturbation; Resonance states}
\narrowtext
 Perturbation for wave functions of quantum systems is
 more complicated problem than that for energy eigenvalues since
 the former deals with functions
 while the latter only deals with a number.
 Even for energy eigenvalues, however, naive perturbations by
 a coupling constant give rise to divergent series in most cases and
 numerous optimization techniques have been studied \cite{div}.
 Variational perturbation method, sometimes called the $\delta$
 expansion, is one of the successful approaches. It
 has been found to reproduce energy eigenvalues of quantum
 anharmonic oscillator (AHO) and double-well potential (DWP)
 with high accuracy \cite{delta}.
 Characteristic of this method is that one introduces artificial
 parameters which are not contained in an original Hamiltonian
 and determines those parameters after calculating a physical
 quantity perturbatively. One of the advantage of the method
 is such flexibility that one can choose different value of the parameter
 depending on the physical quantity one is interested in.
 One can also choose different value depending on the order of
 the perturbation one employs.
 In our previous work, we showed that this advantage could be
 suitably taken to apply the method to wave functions and obtained
 uniformly accurate wave functions of AHO and DWP \cite{opt}.
 
 Recently, it has been shown that a variational perturbation approach
 can be
 also adapted to treat systems which accompany quantum tunneling
 decay processes \cite{tunnel}. In these systems, energy eigenvalues have
 imaginary parts and naive perturbations fail in general.
 In this article, motivated by these studies,
 we attempt to apply our method in \cite{opt} to wave
 functions of a tunneling system. As a simple model, we take
 anharmonic oscillator with negative coupling constant (NAHO).

 We first recall one dimensional AHO to illustrate the variational
 perturbation treatment for wave functions. The Hamiltonian is
\beq
H(\lambda)=\frac{1}{2}\left(-\frac{d^2}{dx^2} + x^2 + \lambda x^4\right).
\label{aho}
\eeq
 One defines a new Hamiltonian $H_{\delta}(\Omega,\lambda)=H_{0}(\Omega)
+\delta H_{I}(\Omega,\lambda)$ with
\beq
H_{0}(\Omega) = \frac{1}{2}\left(-\frac{d^2}{dx^2} + \Omega^2 x^2\right),
\qquad
H_{I}(\Omega,\lambda) = \frac{1}{2}\left( (1-\Omega^2) x^2
 + \lambda x^4\right)
\eeq
 such that $H_{\delta=0}(\Omega,\lambda)=H_{0}(\Omega)$ and
 $H_{\delta=1}(\Omega,\lambda)=H(\lambda)$. At first, one employs
 Rayleigh-Schr\"odinger (RS) perturbation with $\delta$ as an expansion
 parameter. Bases of the expansion are the eigenvectors of $H_{0}(\Omega)$
 which is harmonic oscillator (HO) with a trial frequency $\Omega$.
 The $n$-th order perturbative wave functions thus obtained have the
 following form
\beq
\psi_{k}^{(n)}(x;\Omega) = \sum_{j}\psi_{j}^{0}(x;\Omega)
 \sum_{l=0}^{n} \delta^{l} C_{jk}^{(l)}(\Omega).
\label{pwf}
\eeq
 Next step is to choose the parameter $\Omega$ so as to make
 approximate functions uniformly valid. This can be achieved \cite{opt} by
\beq
\left. \frac{\partial\psi^{(n)}(x;\Omega)}{\partial\Omega}
 \right|_{\delta=1} = 0.
\label{pms}
\eeq
 This condition can be solved for $\Omega$, in principle, as a
 function of $x$ (and $\lambda$). By substituting the resultant $\Omega(x)$
 to (\ref{pwf}), we get optimized wave functions $\psi^{(n)}(x;\Omega(x))$. 

 We can interpret the condition (\ref{pms}) in two different
 ways. First interpretation is a geometrical one. If we regard
 the obtained functions $\psi^{(n)}(x;\Omega)$ as a family
 of curves parametrized by $\Omega$, Eq.(\ref{pms}) is
 nothing but the necessary condition for $\psi^{(n)}(x;\Omega(x))$
 to be an envelope of the family \cite{opt}. The other interpretation
 is the principle of minimum sensitivity (PMS) \cite{stev}; perturbatively
 calculated quantity should not depend on the artificial
 parameter $\Omega$ at $\delta=1$ since the exact one does not 
 depend on $\Omega$ at all. 

 For the ground state wave function, first order variational perturbation
 results in
\beq
\lefteqn{
\psi_{0}^{(1)}(x;\Omega) = \pi^{-1/4}\Omega^{1/4} \exp \left(-
\frac{\Omega}{2} x^2 \right)
}
\nonumber\\
&& \times \left[1+\delta\left\{\frac{1}{8}(1-\Omega^{-2})(2\Omega x^2 -1)
%\right.\right.\nonumber\\ && \left.\left.
-\frac{\lambda}{32}\Omega^{-3}(4\Omega^2 x^4 +12\Omega x^2 -9)\right\}\right].
\label{ground}
\eeq
 The condition (\ref{pms}) applied to (\ref{ground}) reads
\beq
%\lefteqn{
16 x^4 \Omega^5 + 16 x^2 \Omega^4 - 4 (2 \lambda x^6+4 x^4+7) \Omega^3
%}\nonumber\\ &&
 -4 (9 \lambda x^4+4 x^2) \Omega^2 - 2 (33 \lambda x^2-14) \Omega
 + 99 \lambda = 0.
\label{alg}
\eeq
 Since this is a higher degree algebraic equation for $\Omega$,
 one cannot solve it analytically in general. However asymptotic
 behavior of the physical solution at long distance can be
 estimated by Eq.(\ref{alg}) and found to be
\beq
\Omega(x)^2 \to \frac{\lambda}{2} x^2 \qquad (x \to \infty).
\label{asymps}
\eeq
 Therefore, the perturbative wave function improved by the
 condition (\ref{pms}) has its asymptotic form
\beq
\psi_{0}^{(1)}\left(x;\Omega(x)\right) \to \exp \left(-
\sqrt{\frac{\lambda}{8}} |x|^3 \right) \qquad (x \to \infty).
\label{asympw}
\eeq
 The result obtained above has correct $x$ dependence and
 only the numerical factor $8$ deviates a bit from the exact
 value $9$. A noticeable success of the method in this
 case can be appreciated by this correct asymptotic behavior.
 One can never obtain such a behavior by naive RS perturbation
 since perturbative wave functions obtained by RS at any finite
 order result in finite superposition of the wave functions of HO.
 For example, the ground state wave function obtained by naive RS
 perturbation is of the following expression\footnote{Note that this form
 is independent of whether $\lambda>0$ or $\lambda<0$.}
\beq
\psi_{RS_{0}}^{(1)}(x) = \pi^{-1/4} \exp\left(-\frac{1}{2}x^{2}\right)
 \left[1-\frac{\lambda}{32}\left(4x^{4}+12x^{2}-9\right)\right] .
\label{rsw}
\eeq

 In the case of $\lambda > 0$ the physical solutions of
 Eq.(\ref{alg}) are to be restricted to real number but
 this is not the case for $\lambda <0$. The situation is
 observed by the asymptotic solutions (\ref{asymps});
 by Eq.(\ref{asymps}) with $\lambda < 0$ we have
\beq
\Omega(x) \to \pm i \sqrt{\frac{|\lambda|}{8}} |x|,
\label{asymps2}
\eeq
 that is, pure imaginary at infinity. Substituted them into
 Eq.(\ref{ground}) reads
\beq
\psi_{0}^{(1)}\left(x;\Omega(x)\right) \to \exp \left(\mp i \sqrt{
\frac{|\lambda|}{8}} |x|^3\right).
\label{asympw2}
\eeq
 This oscillatory behavior is of course correct one for
 NAHO except for the numerical coefficient.

 From the consideration above, we may expect to get resonance
 states of NAHO perturbatively from the bound-state
 wave functions of HO if we make a novel generalization of our method
 to allow the trial frequency $\Omega$ to take complex values.
 In fact, for the ground state energy of NAHO, it was shown by Karrlein
 and Kleinert \cite{tunnel} that the imaginary part of the energy
 can be extracted precisely by this complex-valued 
 optimization procedure for energy eigenvalues.

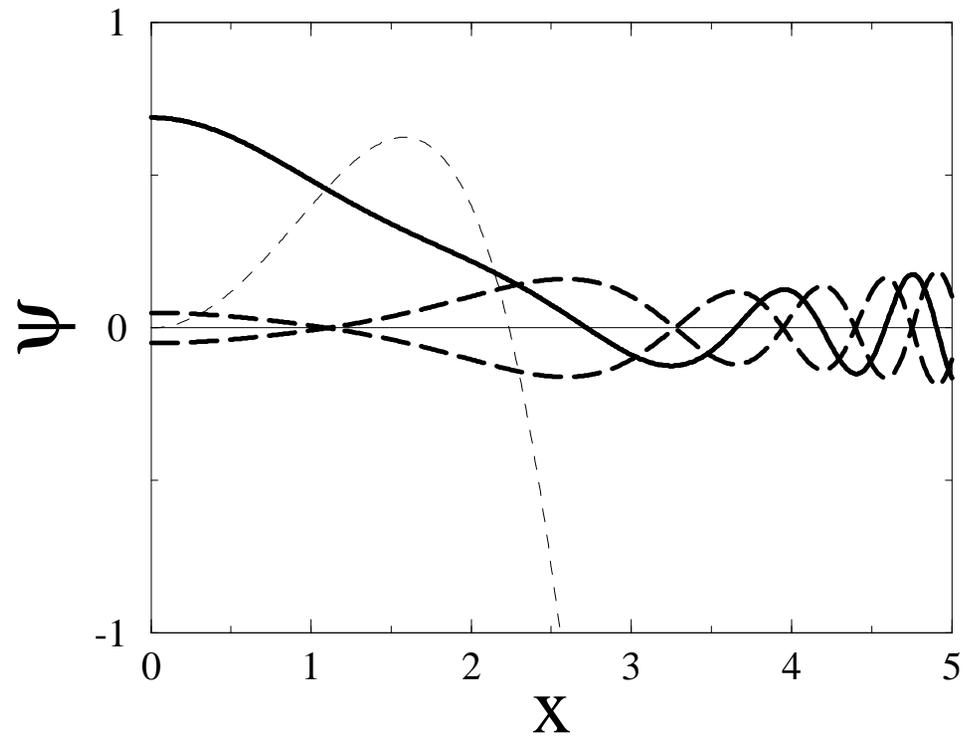
\begin{figure}
%\centerline{\epsfxsize=8.5cm \epsfbox{fig1.ps}}
\caption{The perturbative wave functions $\psi^{(1)}(x,\Omega(x))$ for
 the ground state with $\lambda=-0.2$. The thick solid line
 is their real part and the thick dashed lines are their imaginary
 parts. (See text for the meaning of two lines.)
 The thin dashed line is the potential with $\lambda=-0.2$.
 Relation between types of lines and their meanings is common to all
 the figures in this article.}
\label{fig1}
\end{figure}

 In Fig.\ref{fig1} we show as an example the ground state wave functions
 obtained in this way for $\lambda=-0.2$. There are two wave functions
 which are complex conjugate each other. Thus, there appear two thick
 dashed lines in Fig.\ref{fig1} which correspond to the imaginary parts
 of these two wave functions.
 The real part contributes high probability density in the vicinity
 of the potential valley and decreases as the potential
 barrier becomes high. Outside the barrier, both the real
 part and imaginary part oscillate as Eq.(\ref{asympw2})
 indicates. Therefore, we successfully get qualitative
 behavior of the resonant wave functions of NAHO even at
 the first order perturbation.

 When we try to compare the perturbatively obtained results
 with the exact ones, we encounter two difficulties. Firstly,
 it is difficult to compute numerically an eigenstate
 which oscillates rapidly at boundaries. Secondly,
 the wave functions under consideration are not
 normalizable and therefore it becomes meaningless to
 compare numerical values of the wave functions at
 each point. To avoid these problems, we invoke the
 complex-coordinate method (CCM) \cite{crm}. This tool is often
 used for computing complex eigenvalues for resonance states
 numerically. The essence of CCM is as follows. The ordinary
 Schr\"odinger equations in configuration space
\beq
{\hat H}\psi(x) = E \psi(x)
\eeq
 are defined on the real $x$ axis. Then, one rotates this support
 of $\psi(x)$ in complex $x$ plane as $x \to xe^{i\theta}$.
 One determines the argument of the rotation $\theta$
 such that the rotated wave functions
 $\psi(xe^{i\theta})={\hat U}(\theta) \psi(x)$ tend to $0$
 at infinity, that is,
\beq
{\hat U}(\theta){\hat H}{\hat U}^{-1}(\theta) \psi(xe^{i\theta})
 = E \psi(xe^{i\theta})
\label{ccm}
\eeq
\beq
\psi(xe^{i\theta}) \to 0 \qquad (|x| \to \infty)
\label{bc}
\eeq
 where ${\hat U}(\theta)$ denotes the rotation operator in complex
 $x$ plane.
 Thanks to (\ref{bc}), numerical calculation turns out to be
 easier and the rotated wave functions become normalizable.
 
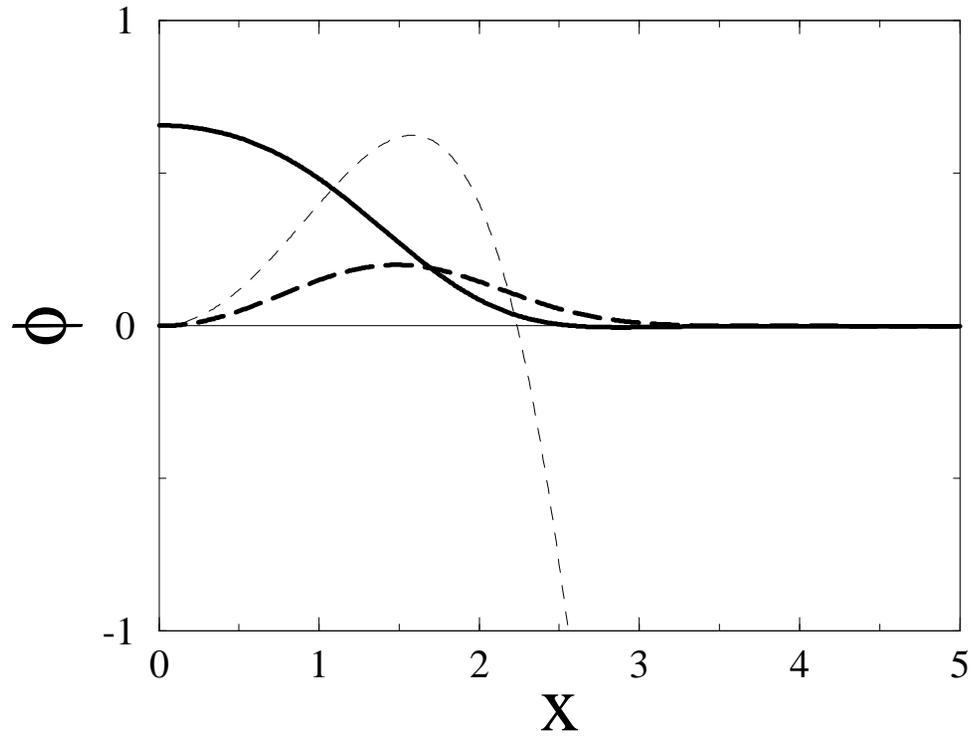
\begin{figure}
%\centerline{\epsfxsize=8.5cm \epsfbox{fig2.ps}}
\caption{The exact wave function $\phi^{\rm exact}(x)=\psi^{\rm exact}
(xe^{i\theta})$ for the ground state with
 $\lambda=-0.2$ calculated by CCM. The rotation argument $\theta$
 equals $-\pi/6$.}
\label{fig2}
\end{figure}

 In order that the wave functions of quartic AHO
 satisfy the boundary condition (\ref{bc}),
 the following inequality must holds \cite{bw}
\beq
\left| (\arg \pm x) + \frac{1}{6} (\arg \lambda) \right| < \frac{\pi}{6}.
\label{bwc}
\eeq
 For NAHO, $\arg\lambda=\pi$, and the inequality (\ref{bwc})
 reads
%\footnote{This condition corresponds to the boundary condition
% for the ordinary wave function defined on the real $x$ axis
% that only in-coming wave exists at infinity.}

\beq
- \frac{\pi}{3} < \theta < 0.
\eeq
 We fix the argument at $\theta=-\pi/6$ for our calculations.
 The normalized ground-state wave function in CCM for $\lambda=-0.2$
 calculated numerically is shown in Fig.\ref{fig2}.

 Finally, we combine perturbation theory with CCM. 
 The procedure is that we rotates Eq.(\ref{pwf}) in
 the complex $x$ plane by the same argument as that for
 the exact one Eq.(\ref{ccm}) ($\theta=-\pi/6$ in our case).
\beq
\psi^{(n)}(x;\Omega) \to \psi^{(n)}(xe^{i\theta};\Omega)
\eeq
 The optimization condition Eq.(\ref{pms}) thus changes in
 the same manner. If the physical solutions of Eq.(\ref{pms})
 are written formally as $\Omega = \Omega (x)$, the solutions
 in the rotated coordinate are $\Omega = \Omega (xe^{i\theta})$.
 Therefore we eventually get the optimized wave functions
 combined with CCM
\beq
\phi_{VP}^{(n)}(x) = \psi^{(n)}(xe^{i\theta};\Omega(xe^{i\theta})).
\label{pwfccm}
\eeq

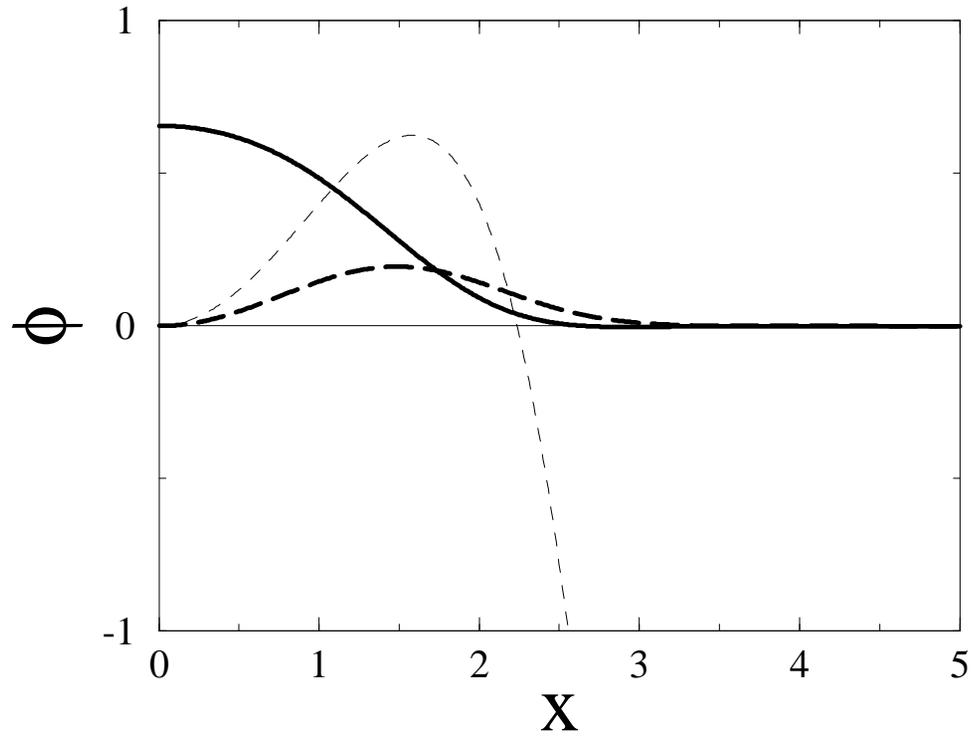
\begin{figure}
%\centerline{\epsfxsize=8.5cm \epsfbox{fig3.ps}}
\caption{The variational perturbation wave function $\phi_{VP}^{(1)}(x)$
 combined with CCM for the ground state with $\lambda=-0.2$.}
\label{fig3}
\end{figure}

 In Fig.\ref{fig3}, we show the 1st-order ground-state wave function
for NAHO with $\lambda=-0.2$ and $\theta=-\pi/6$, obtained by
 Eq.(\ref{pwfccm})\footnote{After we calculate the optimized wave function
 Eq.(\ref{pwfccm}), we make a (global) phase transformation
 so that the phase of the optimized wave function becomes
 the same as that of the exact wave function at the origin.}.
 The comparison of it to the exact one in Fig.\ref{fig2} exhibits excellent
 agreement in the whole of a domain caluculated.
 To make more quantitative comparison,
 (squared) norm defined by
\beq
\Vert \phi^{\rm exact}-\phi^{(1)} \Vert^2 = \int dx |\phi^{\rm exact}(x)
 - \phi^{(1)}(x)|^2
\label{sqn}
\eeq
 are represented in Table \ref{tab1} and the expectation value
 of the (rotated) Hamiltonian are represented in Table \ref{tab2}.
 The agreement is excellent as a first-order perturbation.

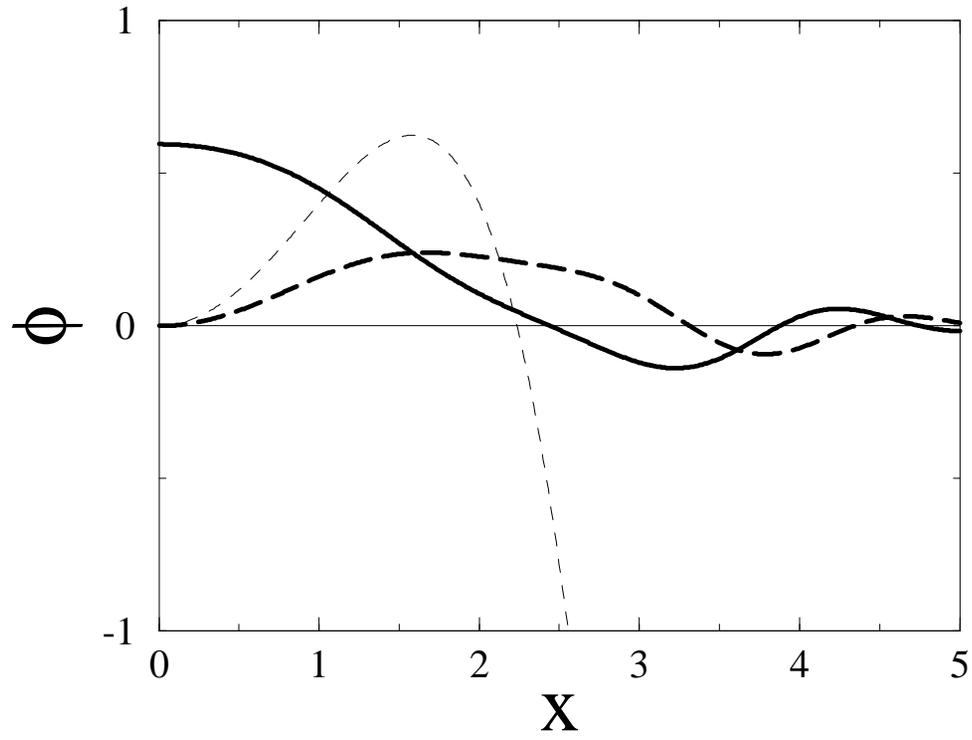
\begin{figure}
%\centerline{\epsfxsize=8.5cm \epsfbox{fig3.ps}}
\caption{The RS perturbation wave function $\phi_{RS}^{(1)}(x)$ combined
 with CCM for the ground state with $\lambda=-0.2$.}
\label{fig4}
\end{figure}

 For the purpose of comparing our variational results with naive
 RS ones, we also show the results obtained from RS perturbation.
 We calculate RS wave functions Eq.(\ref{rsw}) on complex coordinate
\beq
\phi_{RS}^{(0)}(x) = \psi_{RS}^{(0)}(xe^{i\theta})
\eeq
 with $\theta=-\pi/6$. In Fig.\ref{fig4}, $\phi_{RS}^{(0)}$ thus obtained
 with $\lambda=-0.2$ is shown. Squared norm Eq.(\ref{sqn}) and
 the expectation value of the (rotated) Hamiltonian for $\phi_{RS}^{(0)}$
 are also shown in Table \ref{tab1} and \ref{tab2} respectively.

 Some remarks are in order here.\\
\noindent
 1. For $\lambda\leq -0.12$ the physical solutions of Eq.(\ref{alg})
 with real $x$ can be obtained as a complex conjugate pair
 in the whole $x$ domain calculated. They
 correspond to out-going and in-coming states respectively.
 For weak negative coupling $\lambda\geq -0.12$, that is, high potential
 barrier case, there appear domains where the physical solutions
 split into two different real values and one cannot judge which
 solution should be taken. Therefore naive application of the
 optimization condition (\ref{pms}) do not work well in high-barrier
 case. This situation is similar to that for the same optimization
 scheme for eigenvalues of NAHO \cite{tunnel}.

\noindent
 2. In the case of the optimized perturbation combined with CCM,
 the physical solutions of Eq.(\ref{alg}) with complex $x$ can be
 determined uniquely as a complex number since the coefficients
 of the algebraic equation (\ref{alg}) are now complex.
 For $\lambda\leq -0.12$, imaginary parts of those physical solutions are
 all positive in the whole domain calculated.
 For $\lambda\geq -0.12$, on the other hand, imaginary parts of
 the physical solutions turn to be negative in a domain $|x|\leq x_{0}$
 with a certain point $x_{0}$. However, no difficulties
 arise in the method with CCM in contrast to in the case of the method
 with real $x$.

\noindent
 3. From Table \ref{tab1} we can see that the errors for our method
 take its maximum around $\lambda=-0.2$. Even at this point, however,
 the error defined by Eq.(\ref{sqn}) is about 0.0016, which
 indicates that our method is applicable to both the tunnel
 regime and the sliding regime of the coupling constant space\footnote{
 Some authors characterize the tunneling (sliding) regime of the coupling
 constant space as the real part of the ground state energy lower (higher)
 than the potential barrier height and we follow this definition here.
 For our Hamiltonian Eq.(\ref{aho}), the border of the two regime is
 at $\lambda\approx -0.3387$.}. 
 On the contrary, in the case of the optimization for
 eigenvalues of NAHO, it was proved that there exists lower bound
 $\lambda_{0}$ so that optimized perturbation series converge
 to the exact ones for $|\lambda| \geq \lambda_{0}$ \cite{conv}.
 Therefore, it may be said that
 convergence property of the method for eigenfunctions with CCM is different
 from that for eigenvalues.

\noindent
 4. We have also calculated the 2nd-order variational perturbation.
 The results only exhibit slight quantitative improvement from
 the 1st-order ones. This feature indicates rapid convergence
 property of the method although whether perturbative
 wave functions really converge into the exact ones or not is
 of course another problem.

\noindent
 5. In view of both perturbation theory and physical problem,
 Zeeman and Stark effects are interesting problems\footnote{An interesting
 resummation approach for Stark effect is found in Ref.\cite{stark}.}.
 These systems accompany quantum tunneling and thus the method presented
 in this paper will be applicable.
 
 In summary, we have generalized the variational perturbation
 method for wave functions of bound states to treat those of
 resonant states and applied it for NAHO.
 We can obtain qualitative nature of resonance states such as
 oscillatory asymptotic behavior and reproduction
 of complex energy, only by a 1st-order perturbation based on
 the bound states of HO. The method with CCM can overcome the difficulty
 in the method with real $x$ for $\lambda\geq -0.12$ and can give
 uniformly accurate wave functions in both the tunneling and the sliding
 regime.

\section*{Acknowledgements}
 The author thanks T.Kunihiro and T.Hatsuda for helpful
 discussions and comments. The author also thanks H.Suzuki
 for informing their works related to this article.

%\end{thebibliography}

%\newpage

\begin{table}  
\begin{center}
\begin{tabular}{|c|c|c|}
 & $\Vert\phi^{\rm exact}-\phi_{VP}^{(1)}\Vert^2 \quad$
 & $\Vert\phi^{\rm exact}-\phi_{RS}^{(1)}\Vert^2 \quad$ \\
\hline\hline
$\lambda=-0.05$ & $1.29994 \;$E-5 & $1.04105 \;$E-3 \\
\hline
$\lambda=-0.06$ & $3.34356 \;$E-5 & $1.98414 \;$E-3 \\
\hline
$\lambda=-0.07$ & $7.73821 \;$E-5 & $3.38198 \;$E-3 \\
\hline
$\lambda=-0.08$ & $1.59194 \;$E-4 & $5.31350 \;$E-3 \\
\hline
$\lambda=-0.09$ & $2.89919 \;$E-4 & $7.84056 \;$E-3 \\
\hline
$\lambda=-0.1$ & $4.75445 \;$E-4 & $1.10032 \;$E-2 \\
\hline
$\lambda=-0.2$ & $1.56666 \;$E-3 & $7.65252 \;$E-2 \\
\hline
$\lambda=-0.3$ & $1.18456 \;$E-3 & $1.90267 \;$E-1 \\
\hline
$\lambda=-0.4$ & $9.16755 \;$E-4 & $3.32428 \;$E-1 \\
\hline
$\lambda=-0.5$ & $7.57480 \;$E-4 & $4.84690 \;$E-1 \\
\hline
$\lambda=-0.6$ & $6.56376 \;$E-4 & $6.33762 \;$E-1 \\
\hline
$\lambda=-0.7$ & $5.87611 \;$E-4 & $7.72179 \;$E-1 \\
\hline
$\lambda=-0.8$ & $5.38161 \;$E-4 & $8.96876 \;$E-1 \\
\hline
$\lambda=-0.9$ & $5.01007 \;$E-4 & $1.00742 \;$E0 \\
\hline
$\lambda=-1.0$ & $4.72111 \;$E-4 & $1.10469 \;$E0
\end{tabular}
\end{center}
\caption{Squared norm defined by Eq.(19). 
 Since both $\phi^{\rm exact}$ and $\phi^{(1)}$ are normalized to unity,
 the squared norm takes a value in the interval [0,2].}
\label{tab1}
\end{table}

\newpage

\begin{table}  
\begin{center}
\begin{tabular}{|c|c|c|c|}
 & $\langle H \rangle^{\rm exact}$
 & $\langle H \rangle_{VP}^{(1)}$ & $\langle H \rangle_{RS}^{(1)}$ \\
\hline\hline
$\lambda=-0.05$ & $0.47912 + i \; 0.00001$ & $0.47971 + i \; 0.00036$
 & $0.50328 - i \; 0.00369$ \\
\hline
$\lambda=-0.06$ & $0.47416 + i \; 0.00006$ & $0.47519 + i \; 0.00061$
 & $0.51141 - i \; 0.00176$ \\
\hline
$\lambda=-0.07$ & $0.46879 + i \; 0.00026$ & $0.47050 + i \; 0.00098$
 & $0.52287 + i \; 0.00248$ \\
\hline
$\lambda=-0.08$ & $0.46297 + i \; 0.00077$ & $0.46564 + i \; 0.00150$
 & $0.53799 + i \; 0.00961$ \\
\hline
$\lambda=-0.09$ & $0.45677 + i \; 0.00176$ & $0.46062 + i \; 0.00219$
 & $0.55712 + i \; 0.02025$ \\
\hline
$\lambda=-0.1$ & $0.45034 + i \; 0.00335$ & $0.45551 + i \; 0.00308$
 & $0.58058 + i \; 0.03497$ \\
\hline
$\lambda=-0.2$ & $0.39744 + i \; 0.04471$ & $0.39512 + i \; 0.03694$
 & $1.11140 + i \; 0.52301$ \\
\hline
$\lambda=-0.3$ & $0.37377 + i \; 0.09506$ & $0.36928 + i \; 0.09092$
 & $2.28179 + i \; 1.88598$ \\
\hline
$\lambda=-0.4$ & $0.36441 + i \; 0.13868$ & $0.35999 + i \; 0.13638$
 & $4.06817 + i \; 4.22403$ \\
\hline
$\lambda=-0.5$ & $0.36146 + i \; 0.17576$ & $0.35740 + i \; 0.17444$
 & $6.30746 + i \; 7.37246$ \\
\hline
$\lambda=-0.6$ & $0.36169 + i \; 0.20780$ & $0.35798 + i \; 0.20704$
 & $8.81116 + i \; 11.06949$ \\
\hline
$\lambda=-0.7$ & $0.36363 + i \; 0.23600$ & $0.36022 + i \; 0.23560$
 & $11.43170 + i \; 15.07917$ \\
\hline
$\lambda=-0.8$ & $0.36655 + i \; 0.26122$ & $0.36339 + i \; 0.26106$
 & $14.07565 + i \; 19.23405$ \\
\hline
$\lambda=-0.9$ & $0.37004 + i \; 0.28407$ & $0.36710 + i \; 0.28408$
 & $16.69183 + i \; 23.42979$ \\
\hline
$\lambda=-1.0$ & $0.37387 + i \; 0.30499$ & $0.37111 + i \; 0.30512$
 & $19.25816 + i \; 27.61088$
\end{tabular}
\end{center}
\caption{Expectation value of (rotated) Hamiltonian with
 respect to the exact wave function $\langle H \rangle^{\rm exact}$
 and that with respect to the perturbative one $\langle H \rangle^{(1)}$.}
% For all $\lambda$ calculated, $\langle H \rangle^{\rm exact}$ agree to
% the exact energy eigenvalues up to the order $10^{-7}$.}
\label{tab2}
\end{table}

\end{document}